# Very large magneto-impedance and its scaling behavior in amorphous $Fe_{73.5}Nb_3Cu_1Si_{13.5}B_9$ ribbon

B. Kaviraj[*]

LGEP/SPEE Labs ; CNRS UMR 8507 Supelec ; Univ. Pierre et Marie Curie-P6 ; Univ. Paris Sud-P11, 11 Rue Joliot-Curie, Plateau de Moulon, 91192 Gif sur Yvette, France.

Tel : +0033(0)6-72-58-24-76

Email: bhaskar.kaviraj@lgep.supelec.fr

S.K. Ghatak

Magnetism Laboratory, Department of Physics & Meteorology, Indian Institute of Technology, Kharagpur 721302, India. Tel: +91-3222-283818  Fax: +91-3222-255303

Email: skghatak@phy.iitkgp.ernet.in

[*] Corresponding Author

# Abstract

Magneto-impedance (MI) effects have been observed for amorphous $Fe_{73.5}Nb_3Cu_1Si_{13.5}B_9$ ribbon which has been excited by an a.c. magnetic field parallel to the length of the ribbon. Maximum relative change in MI as large as -99% was observed in the as-cast state of the ribbon. The relative change in MI, $\delta Z = [Z(H) - Z(0)]/Z(0)$ when plotted against scaled field $H/H_{1/2}$ was found to be nearly frequency independent; $H_{1/2}$ is the field where $\delta Z$ reduced to half its maximum. A phenomenological formula for magneto-impedance, $Z(H)$, in a ferromagnetic material, is proposed based on Pade' approximant to describe the scaled behavior of $\delta Z$.

**Introduction**

The magneto-impedance effect[1,2] – a large change in impedance of a soft ferromagnetic conductor in presence of a d.c. magnetic field is at the basis of development of advanced magnetic sensors with high sensitivity and high spatial resolution[3-5]. Interest in GMI was initiated in the early nineties when Panina *et al.*[2] and Beach *et al.*[6] reported a very large effect in amorphous ferromagnetic FeCoSiB wires at small magnetic fields and at relatively low frequencies. Since then, the GMI effect has been investigated in a variety of Fe- and Co-based amorphous ribbons,[7-11] films,[12-14], wires and polycrystalline oxides[15-19]. It is generally understood that the change of impedance is a direct consequence of the dependence of the skin effect on the relative magnetic permeability[20-23]. In the domain of low frequency, the MI effect is considered as a classical phenomenon and the explanation is based on solution of Maxwell equations coupled with linear magnetization dynamics. The impedance 'Z' for a magnetic ribbon with thickness '2d' excited by an a.c current through a coil mounted on it can be expressed in terms of complex wave vector 'k' as:

$$Z = -jX_0 \mu_e \left[ \frac{tanh(kd)}{kd} \right] \qquad (1)$$

where '$X_0$' is the reactance of empty sample coil, $\mu_e = \mu_e' + j\mu_e''$ is the effective permeability of the material. The wave vector $k$ is given as $k = \left( \frac{1+j}{\delta_m} \right)$ where $\delta_m = \left[ \frac{2}{\omega \mu_0 \mu_e \sigma} \right]^{1/2}$ is the skin depth,

$\mu_0$ = permeability of free space and $\sigma$ = d.c conductivity of the material. The impedance of a conductor is thus governed by the skin penetration depth $\delta_m$. The behavior of permeability with the bias field changes the skin depth and hence the impedance of the sample. Note that in equation (1), the d.c and a.c fields are parallel to the length of the ribbon. Hitherto, most studies on GMI effects were concentrated for field configurations where the d.c field (along the length of sample) is perpendicular to the a.c. field. In this configuration, almost negligible MI effects were observed for $Fe_{73.5}Nb_3Cu_1Si_{13.5}B_9$ thin films[24] and ribbons[25] although the behavior improved after transverse field annealing of the films and nanocrystallization of ribbons.

In this paper, we report very large MI effects in the amorphous (as-cast) state of $Fe_{73.5}Nb_3Cu_1Si_{13.5}B_9$ ribbon. The ribbon was excited by passing an a.c current through a signal coil that is wound across the ribbon and the voltage (hence the impedance) being measured across the coil itself. This is in contrast to the usual convention where the voltage is usually measured across the sample ends. This is basically a non-contact method for measuring the impedance of a metallic system. The MI data for different frequencies collapse into a single curve when the relative change in MI denoted by $\delta Z = [Z(H) - Z(0)]/Z(0)$ is plotted against the scaled field $H/H_{1/2}$, $H_{1/2}$ is the field where δZ reduces to half its maximum. A phenomenological formula for impedance is proposed to describe the field dependence and collapse of MI data at different frequencies.

**Experiments**

The experiments were carried out with the samples cut from amorphous $Fe_{73.5}Nb_3Cu_1Si_{13.5}B_9$ ribbon prepared using conventional single roller rapid quenching in vacuum. The sample (30μm thick and 4mm wide) with 20mm in length was placed symmetrically within a 50-turns small coil of rectangular geometry such that the exciting a.c field was always along the length of the ribbon. The length of the sample was larger than that of the coil. The real (R) and the imaginary (X) components of impedance Z=R+jX, of the coil with and without the sample were measured using an Agilent Impedance Analyzer (Model-4294A). The impedance of the empty coil was subtracted and only the impedance of the sample was taken into account. The excitation current amplitude was fixed at 10mA and the estimated a.c excitation field was about 140A/m. The response of the sample around this value of exciting field was found to be linear. The applied d.c field was in the plane and along the length of the sample and also along the direction of a.c field.

**Results and Discussions**

The frequency dependence of real (R) and imaginary (X) components of impedance of the amorphous sample in the presence of different bias fields is displayed in Fig. 1. R increases monotonically with frequency. While at low frequencies X exhibits a monotonic increase, at higher frequencies it reaches a maximum and then falls off. The application of a bias field suppresses both the components of impedance. The maximum value of X shifts to higher frequencies with the increase of bias fields. One of the striking features is the strong resistive change at higher frequencies. Similar results were reported by Machado[26] et al. for Co-Fe-Si-B ribbons. For a Fe-Co-Si-B ribbon[27], Beach et al. reported that R had a positive curvature while X showed a negative curvature versus frequency below 3MHz, indicating that R would increase more and more sharply while X would tend to saturate or decrease.

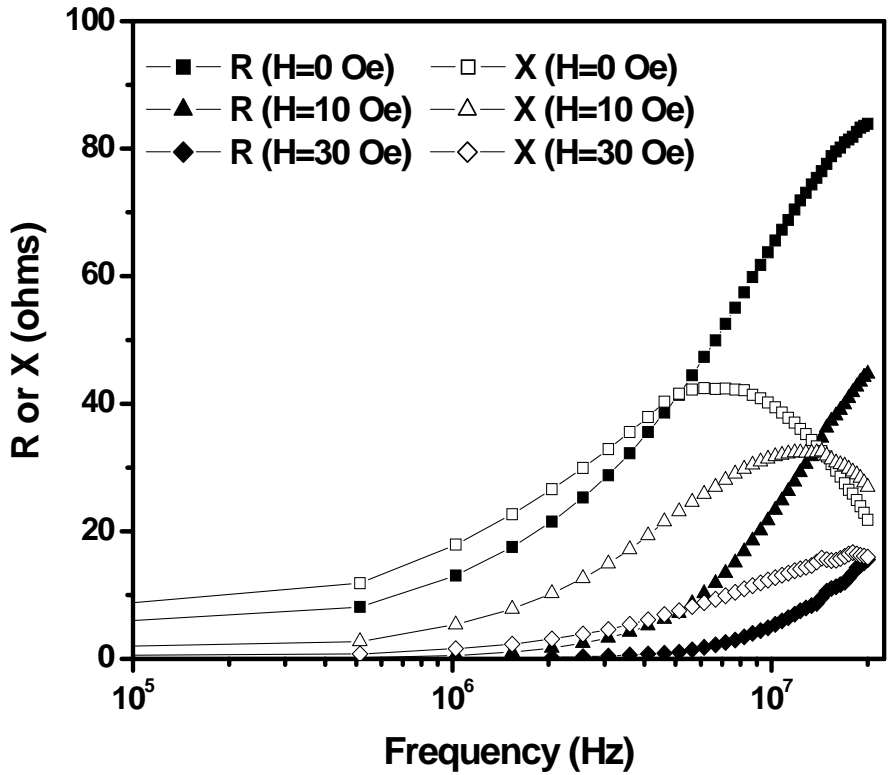

Fig. 1. Frequency dependence of resistive (R) and reactive (X) components of impedance at different bias fields applied along the direction of a.c field for the amorphous $Fe_{73.5}Nb_3Cu_1Si_{13.5}B_9$ ribbon.

In Fig. 2, we depict the field dependence of relative change in real ($\delta R = [R(H) - R(0)/R(0)]\%$) and imaginary ($\delta X = [X(H) - X(0)/X(0)]\%$) components of magneto-impedance at different excitation frequencies. Maximum changes as large as -99% have been observed in both the real and imaginary components of MI and at low frequencies ~ 100KHz.

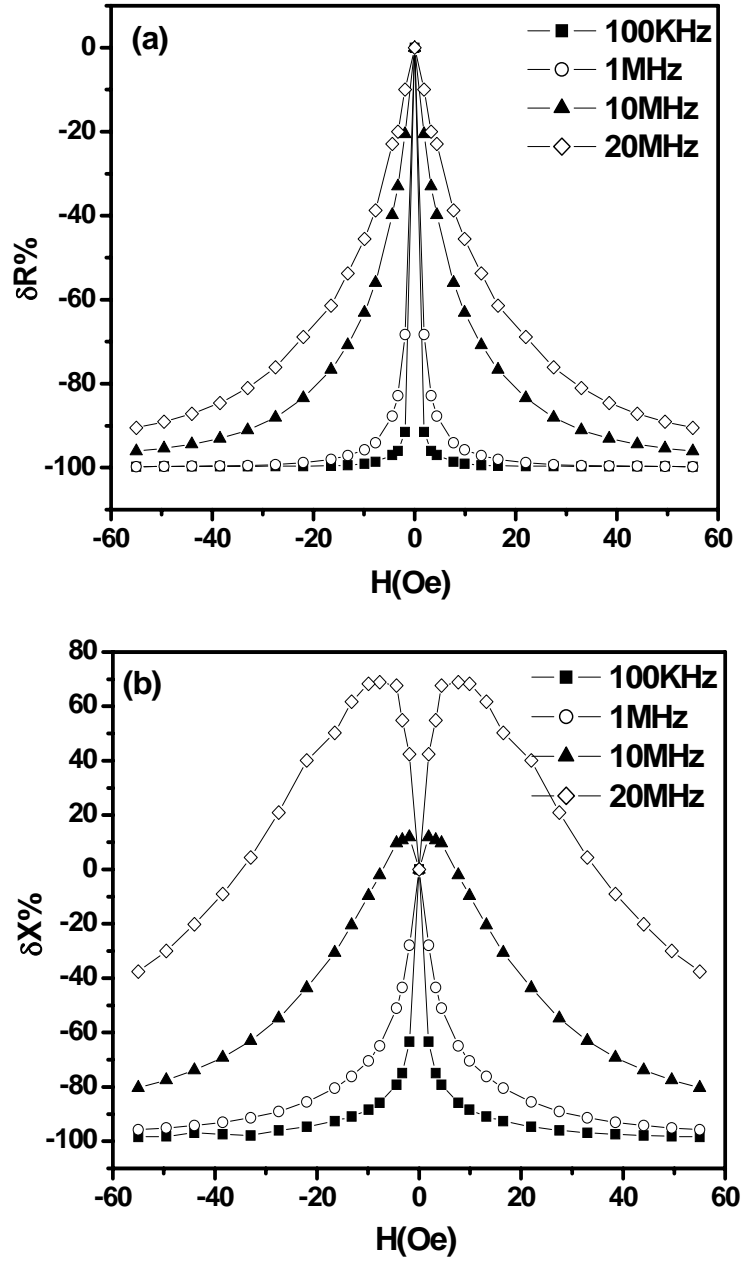

Fig. 2. Field dependence of relative change in real $\delta R = [R(H) - R(0)/R(0)]\%$ and imaginary $\delta X = [X(H) - X(0)/X(0)]\%$ components of MI at different excitation frequencies for the amorphous $Fe_{73.5}Nb_3Cu_1Si_{13.5}B_9$ ribbon. The applied magnetic field was in the plane of ribbon and along the direction of exciting a.c field.

Such a large change in MI is related to large reduction of magnetic response in presence of longitudinal d.c. field. The maximum changes in δR and δX reduces with the increase of frequency. This is attributed to the decrease in permeability at higher frequencies[25]. At frequencies above 1MHz, δX becomes positive at low fields and exhibit peaks. The peak-height of this positive magneto-reactance increases from 10% at 10MHz to 70% at 20MHz. The field $H_k$ where maximum in δX appears also increases from 1.8Oe at 10MHz to 7Oe at 20MHz. Such peaks in the MI spectra were also obtained by Sommer *et al.* for $Fe_{73.5}Nb_3Cu_1Si_{13.5}B_9$ thin films[24] and Chen *et al.* for nanocrystalline $Fe_{73.5}Nb_3Cu_1Si_{13.5}B_9$ ribbons[25]. But the intensities of the peaks as well as the maximum relative changes in MI were much smaller compared to present observations in as-cast state of $Fe_{73.5}Nb_3Cu_1Si_{13.5}B_9$ ribbon.

In our experimental condition, the electric and magnetic fields of electromagnetic (e.m) excitation are perpendicular and parallel to the ribbon length respectively. This is in contrast to the usual measuring situation where the excitation current flows along the length of the ribbon and hence the field directions are interchanged. Assuming the magnetic field along z-direction (which is also the long axis of the ribbon) and electric field $\vec{e}_y$ along y-direction the induced voltage, $V_s$, across the signal coil around the sample is given by[28] $V_s = \int \vec{e}_y \cdot d\vec{l} = ZI_0$ where $I_0$ is the amplitude of a.c current in the coil and Z is given by (1). It follows from Eq. (1) that the magneto-impedance of the material is determined by magnetic response $\mu_e$. In the absence of d.c field, the response to a.c. excitation is very large for soft ferromagnetic system (low magnetic anisotropy) and this means large permeability. In the limit $\delta_m \ll d$, (1) can be approximated as $Z \approx -(j-1)X_0 \mu_e \left(\delta_m/d\right)$. So the impedance varies as $\mu_e^{1/2}$ and is large for material with large $\mu_e$. The other limit can be achieved in presence of large biasing field (H >> h) that reorients magnetization along H and the a.c. magnetization parallel to h is drastically reduced. This in turn increases $\delta_m$ and $Z \approx -jX_0 \mu_e$. Here $\mu_e$ is the permeability in presence of large biasing field and is reduced by an order of magnitude compared to that at H= 0. The large negative MI is therefore the result of additional screening of e.m field in a magnetic metal. The additional screening current depends on the magnitude and rate of variation of a.c. magnetization and hence MI increases with frequency of exciting field. The field dependence of relative change in total impedance $\delta Z = [Z(H) - Z(0)]/Z(0)$ % at various frequencies is depicted in Fig. 3.

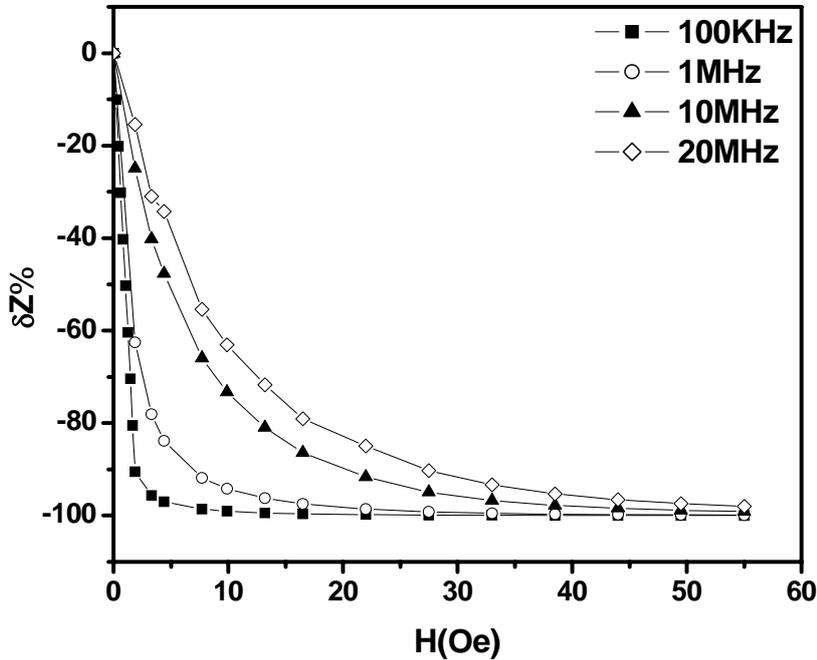

Fig. 3. Field dependence of relative change in total impedance $\delta Z = \frac{(Z(H) - Z(0))}{Z(0)}\%$ at different excitation frequencies for amorphous $Fe_{73.5}Nb_3Cu_1Si_{13.5}B_9$ ribbon.

For comparison, MI measurements have also been performed with the conventional four-probe method in which the directions of excitation field and d.c bias field are mutually perpendicular. The MI curves under this condition are presented in Fig. 4 at 5MHz frequency. As expected and already observed in the previous investigations [24,25], the MI effect in the amorphous $Fe_{73.5}Nb_3Cu_1Si_{13.5}B_9$ ribbon is very small and extends only to a few percents at maximum fields H~ 60Oe.

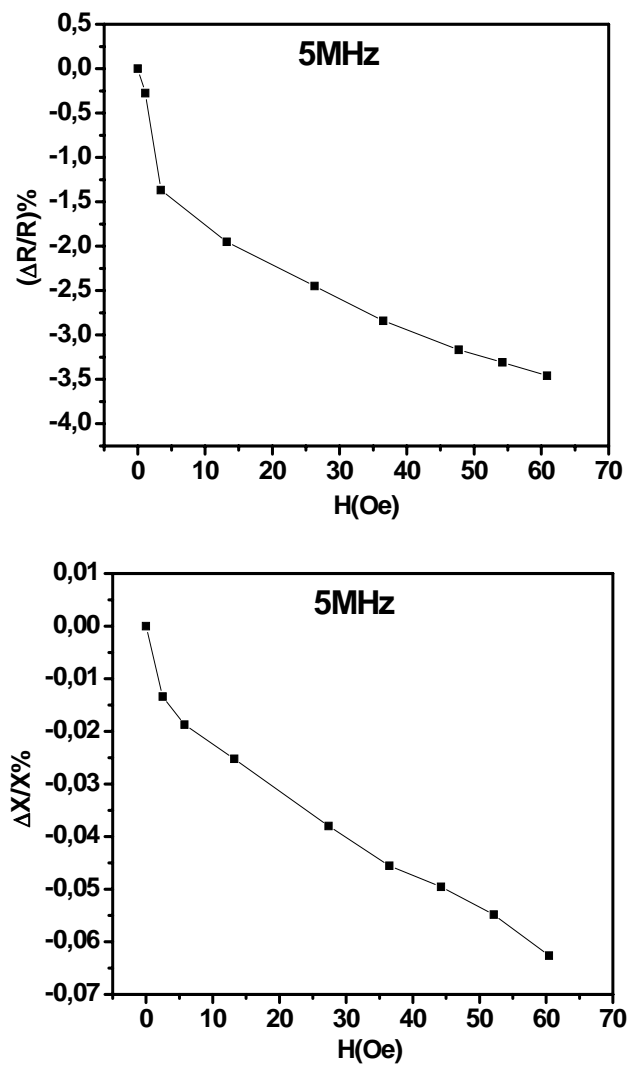

Fig. 4. MI curves for amorphous $Fe_{73.5}Nb_3Cu_1Si_{13.5}B_9$ ribbon at 5MHz frequency. Note in this case that the excitation a.c field and d.c bias field are mutually perpendicular.

In order to look for a more general behavior of magneto-impedance, $\delta Z = [Z(H) - Z(0)]/Z(0)$ (from Fig. 3) has been plotted as functions of $H/H_{1/2}$ at different frequencies. This is shown in Fig. 5. $H_{1/2}$ is the field where $\delta Z$ reduces to half its maximum value and is different for different frequencies. It is remarkable that all the data points collapse into a single curve indicating nearly frequency independent relationship between the reduced quantities $H/H_{1/2}$ and $\delta Z$.

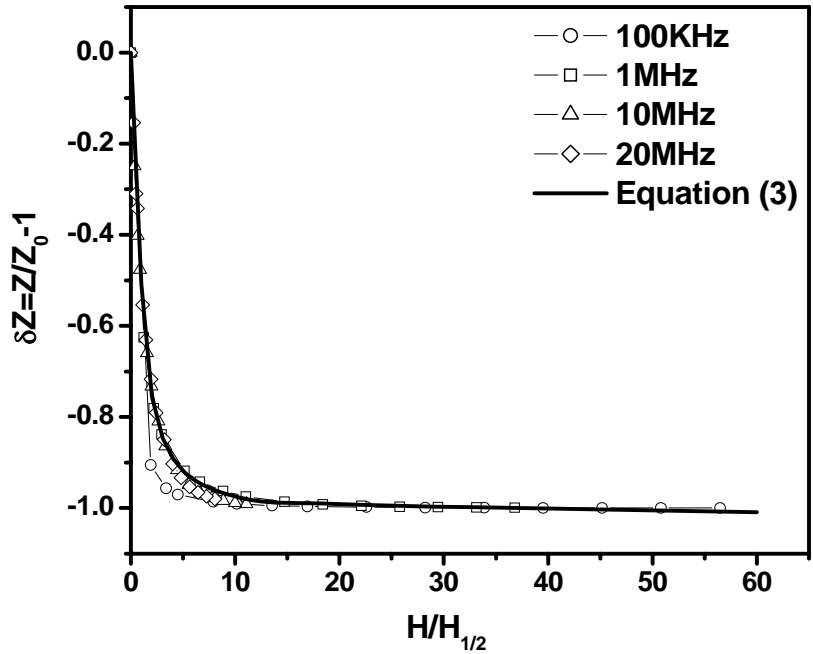

Fig. 5. The relative change in impedance as functions of scaled field $H/H_{1/2}$. The experimental points for different frequencies superposes with equation (3) with $H_0=0.5$ and $H_2=0.51$.

In order to describe the reduced graph, a phenomenological formula is derived below using the results of dependence of Z on H and the Pade approximant. It is observed from the high field region data of Z that dZ/dH can be approximated as $|dZ/dH| \propto H^{-2}$. On the other hand, at very low H, dZ/dH < 0. Considering these results and following the Pade approximation we write dZ/dH as a ratio of two polynomials:

$$\left(\frac{dZ}{dH}\right) = \frac{-A - BH}{C + DH^3} \qquad (2)$$

where A,B,C,D are parameters independent of H. Integrating equation (2) and with the boundary condition that magneto impedance $Z \to 0$ as $H \to \infty$, the equation for Z takes the form:

$$\frac{Z}{Z_0} - 1 = dZ = \frac{3}{2\pi}\left(\frac{\pi}{2} - \tan^{-1}\left(\frac{2H_r - 1}{\sqrt{3}}\right)\right) + \left(\frac{1 + H_{0r}}{1 - H_{0r}}\right)\frac{\sqrt{3}}{4\pi}\log\frac{(H_r + 1)^2}{H_r^2 - H_r + 1} - 1 \qquad (3)$$

where reduced fields $H_r = H/H_2$ and $H_{0r} = H_0/H_2$ with $H_0 = A/B$ and $H_2 = (C/D)^{1/3}$ and $Z_0$ is value of Z at $H=0$. The expression (3) relates the dependence of relative change in Z (denoted by dZ) upon parameters H, $H_0$ and $H_2$. Two representative plots exhibiting the dependence of dZ on field parameter H for different values of $H_0$ and $H_2$ are given in Fig. 6(a) and 6(b).

The curves in Fig. 6 show that the impedance dZ decreases monotonically with field H for low values of $H_0$ (curves with $H_0 = 0.5$ and -0.3 are depicted). The sharpness and extent of decrease in dZ depends on the values of $H_2$. For small values of $H_2$, the decrease is nearly 100% and dZ sharply falls as H increases. With the increase in $H_2$, a slower variation of dZ is found in both the cases. With higher value of $H_0$, the qualitative behavior of dZ is same. For $H_0 < 0$, instead of monotonic decrease, dZ increases as H increases from zero and passes through a maximum for small values of $H_2$ (Fig. 6(b)).

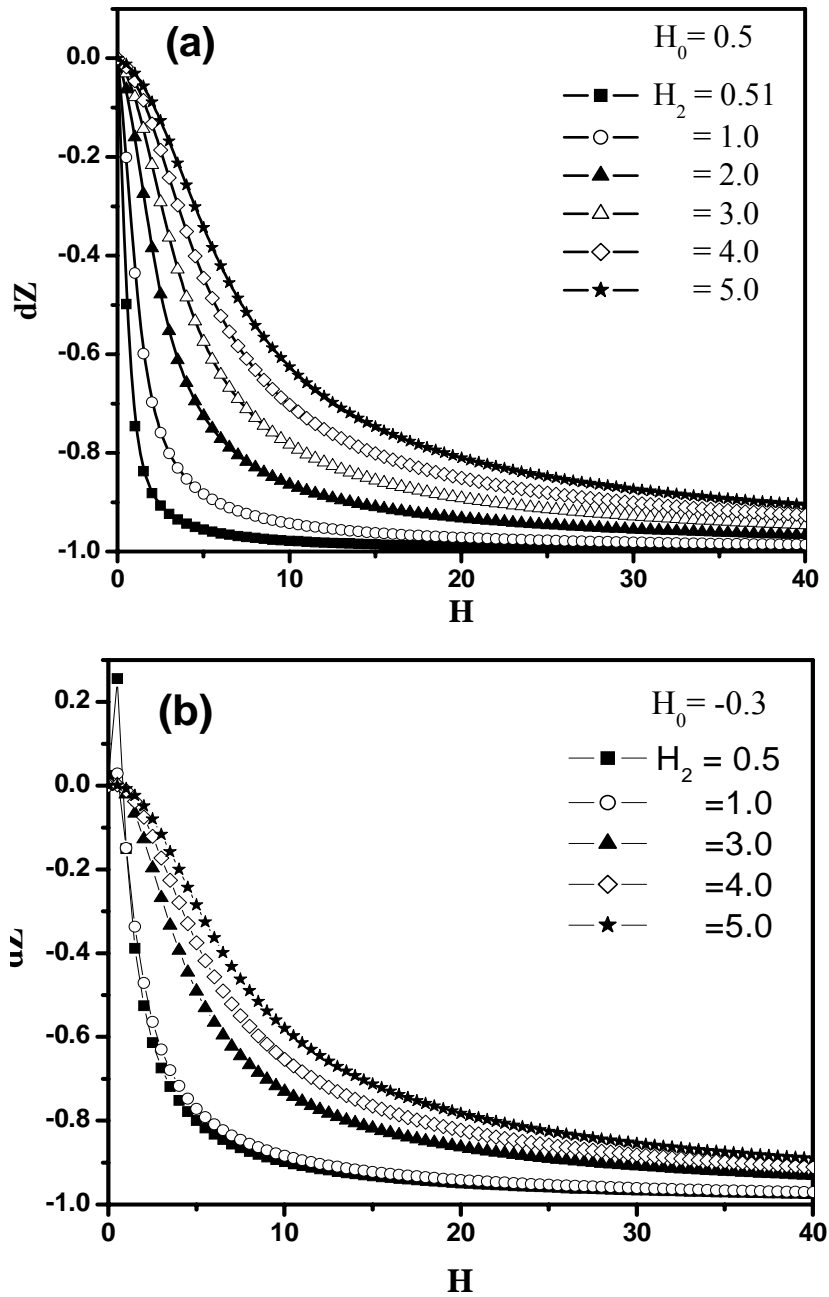

Fig. 6. Plot of dZ as a function H for $H_0 = 0.5$ (a) and $H_0 = -0.3$ (b) according to equation (3) and at various values of $H_2$ as denoted in the plots.

However, this maximum disappears when $H_2 \gg H_0$. As the experimental data is closer to the situation with lower values of $H_0$, we examine the results of Fig. 6(a) in a different way. The results are re-plotted in terms of scaled variables $H/H_{1/2}$ in Fig. 7.

Here $H_{1/2}$ is the field where dZ becomes half of its maximum value. The results of dZ for different $H_2$ collapse into a single curve. As $H_{1/2}$ is higher for larger $H_2$, the curve for higher $H_2$ shifts to lower values of $H/H_{1/2}$ compared to the curve for lower $H_2$. The value of $H_0$ is taken to be smaller than $H_2$ in order to reproduce the observed small slope of dZ near zero field.

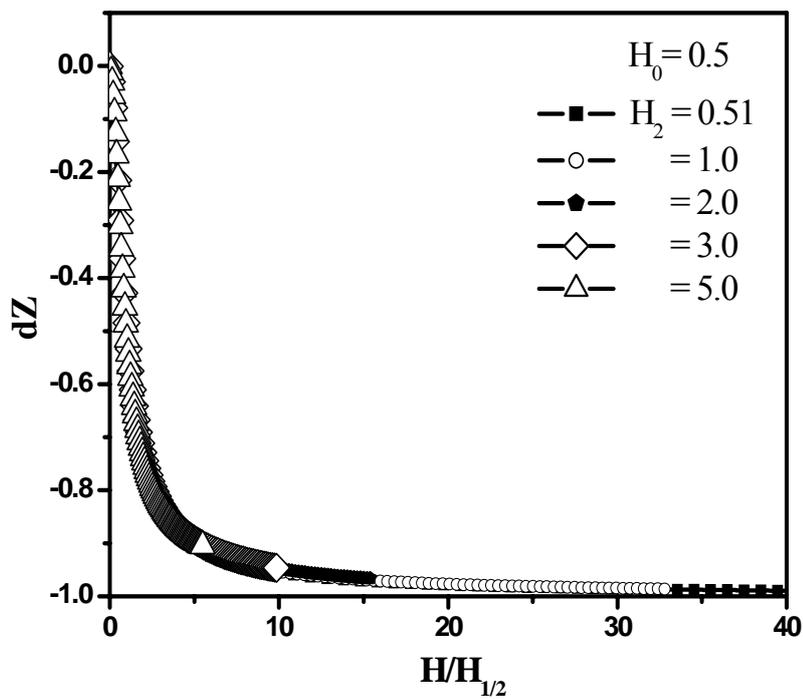

Fig. 7. Re-plot of data in Fig. 5(a) in terms of reduced field $H/H_{1/2}$.

The experimental result ($\delta Z$ vs $H/H_{1/2}$) for the ribbon at different frequencies is almost superposed with the results of equation (3) (Fig. 5). Fig. 5 shows an excellent agreement between the phenomenological equation (3) and the experiment and it demonstrates that equation (3) can be used to describe the scaled behavior of relative Z with respect to the scaled field $H/H_{1/2}$.

**Conclusions**

In conclusion, very large MI effects have been observed in the as-cast state of amorphous $Fe_{73.5}Nb_3Cu_1Si_{13.5}B_9$ ribbon. The large MI is attributed to the large reduction of magnetic response (a.c. permeability) in presence of d.c field (H) much larger than the a.c. field amplitude. For parallel configuration of both fields and large H, the system becomes mono-domain in nature and the total magnetization is pinned and thereby induced magnetization due to a.c. excitation is almost reduced to zero. On the other hand for perpendicular configuration of fields there will be higher induced magnetization for all H and this in turn produces smaller change in Z. The demagnetizing factor, which is inversely proportional to the induced permeability at low d.c bias fields, is also appreciable when the exciting a.c field is perpendicular to the d.c field and along the width of the ribbon. Normally the widths of the ribbons are smaller than their length and this account for low values of induced permeability, and thus, a smaller change in Z in the perpendicular field configuration.

An interesting result is obtained when the relative change in Z is plotted as a function of H scaled to $H/H_{1/2}$ when all the MI curves for different frequencies collapses into a single curve. This shows that the dependence of normalized MI on scaled field $H/H_{1/2}$ is nearly frequency independent. A phenomenological model based on Pade approximant is presented to describe the scaled behavior of δZ where we observe the experimental data to be very well supported by the theoretical model.

# References

[1] L.V. Panina, K. Mohri, K. Bushida and M. Noda, J. Appl. Phys. **76**, 6198 (1994)

[2] L.V. Panina and K. Mohri, Appl. Phys. Lett. **65,** 1189 (1994)

[3] T. Kitoh, K. Mohri and T. Uchiyama, IEEE Trans Magn. **31**, 3137 (1995)

[4] K. Mohri, T. Uchiyama and L.V. Panina, Sens. Actuators A **59**, 1 (1997)

[5] M. Tewes, M. Lohndorf, A. Ludwig, and E. Quandt, Tech. Mess. **68**, 292 (2001)

[6] R. S. Beach and E. Berkowitz, Appl. Phys. Lett. **65**, 3652 (1994)

[7] S. U. Jen and Y. D. Chao, J. Non-Cryst. Solids **207**, 612 (1996)

[8] H. Chiriac, F. Vinai, T. A. Ovari, C. S. Marinescu, F. Barariu, and P. Tiberto, Mater. Sci. Eng., A **226**, 646 (1997)

[9] M. Vazquez, G. V. Kurlyandskaya, J. L. Munoz, A. Hernando, N. V. Dmitrieva, V. A. Lukshina, and A. P. Potapov, J. Phys. I **8**, 143 (1998).

[10] F. Amalou and M. A. M. Gijs, J. Appl. Phys. **90**, 3466 (2001)

[11] E. E. Shalyguina, M. A. Komarova, V. V. Molokanov, C. O. Kim, C. Kim, and Y. Rheem, J. Magn. Magn. Mater. **258**, 174 (2003)

[12] S. Q. Xiao, Y. H. Liu, L. Zhang, C. Chen, J. X. Lou, S. X. Zhou, and G. D. Liu, J. Phys.: Condens. Matter **10**, 3651 (1998)

[13] S. Q. Xiao, Y. H. Liu, S. S. Yan, Y. Y. Dai, L. Zhang, and L. M. Mei, Acta Phys. Sin. **48**, S187 (1999).

[14] J. Q. Yu, Y. Zhou, B. C. Cai, and D. Xu, J. Magn. Magn. Mater. **213**, 32 (2000)

[15] M. Vazquez, M. Knobel, M. L. Sanchez, R. Valenzuela, and A. P. Zhukov, Sens. Actuators A **59**, 20 (1997)

[16] D. X. Chen, J. L. Munoz, A. Hernando, and M. Vazquez, Phys. Rev. B **57**, 10699 (1998)

[17] M. R. Britel, D. Menard, P. Ciureanu, A. Yelon, M. Rouabhi, R. W. Cochrane, C. Akyel, and J. Gauthier, J. Appl. Phys. **85**, 5456 (1999)

[18] K. Mandal, S. Puerta, M. Vazquez, and A. Hernando, Phys. Rev. B **62**, 6598 (2000)

[19] C. Gomez-Polo, M. Vazquez, and M. Knobel, Appl. Phys. Lett. **78**, 246 (2001); V. V. Srinivasu, S. E. Lofland, and S. M. Bhagat, J. Appl. Phys. **83**, 2866 (1998); V. V. Srinivasu, S. E. Lofland, S. M. Bhagat, K. Ghosh, and S. D. Tyagi, J. Appl. Phys. **86**, 1067 (1999)

[20] K. S. Byon, S. C. Yu, and C. G. Kim, J. Appl. Phys. **89**, 7218 (2001)

[21] K. J. Jang, C. G. Kim, S. S. Yoon, and S. C. Yu, Mater. Sci. Eng., A **304**, 1034 (2001)

[22] W. S. Cho, H. Lee, S. W. Lee, and C. O. Kim, IEEE Trans. Magn. **36**, 3442 (2000)

[23] J. F. Hu, L. S. Zhang, H. W. Qin, Y. Z. Wang, Z. X. Wang, and S. X. Zhou, J. Phys. D **33**, L45 (2000)

[24] R.L. Sommer and C.L. Chien, Appl. Phys. Lett. **67**, 3346 (1995)

[25] C. Chen, K.Z. Luan, Y.H. Liu, L.M. Mei, H.Q. Guo, B.G. Shen and J.G. Zhao, Phys. Rev. B **54** 6092 (1999)

[26] F.L.A. Machado, C.S. Martins and S.M. Rezende, Phys. Rev. B **51**, 3926 (1995)

[27] R.S. Beach and A.E. Berkowitz, J. Appl. Phys. **76**, 6209 (1994)

[28] D.P. Makhnovskiy, L.V. Panina and D.J. Mapps, Phys. Rev. B, **63**, 144424 (2001)

**Figure Captions**

Fig. 1. Frequency dependence of resistive (R) and reactive (X) components of impedance at different bias fields applied along the direction of a.c field for the amorphous $Fe_{73.5}Nb_3Cu_1Si_{13.5}B_9$ ribbon.

Fig. 2. Field dependence of relative change in real $\delta R = [R(H) - R(0)/R(0)]\%$ and imaginary $\delta X = [X(H) - X(0)/X(0)]\%$ components of MI at different excitation frequencies for the amorphous $Fe_{73.5}Nb_3Cu_1Si_{13.5}B_9$ ribbon. The applied magnetic field was in the plane of ribbon and along the direction of exciting a.c field.

Fig. 3. Field dependence of relative change in total impedance $\delta Z = \dfrac{(Z(H) - Z(0))}{Z(0)}\%$ at different excitation frequencies for amorphous $Fe_{73.5}Nb_3Cu_1Si_{13.5}B_9$ ribbon.

Fig. 4. MI curves for amorphous $Fe_{73.5}Nb_3Cu_1Si_{13.5}B_9$ ribbon at 5MHz frequency. Note in this case that the excitation a.c field and d.c bias field are mutually perpendicular.

Fig. 5. The relative change in impedance as functions of scaled field $H/H_{1/2}$. The experimental points for different frequencies superposes with equation (3) with $H_0=0.5$ and $H_2=0.51$.

Fig. 6. Plot of dZ as a function H for $H_0 = 0.5$ (a) and $H_0 = -0.3$ (b) according to equation (3) and at various values of $H_2$ as denoted in the plots.

Fig. 7. Re-plot of data in Fig. 6(a) in terms of reduced field $H/H_{1/2}$.